\documentclass[fleqn,usenatbib]{mnras}

\usepackage{newtxtext,newtxmath}

\usepackage[T1]{fontenc}
\usepackage{ae,aecompl}

\usepackage{graphicx}	
\usepackage{amsmath}	
\usepackage{amssymb}	
\usepackage{wrapfig}
\usepackage{lscape}
\usepackage{rotating}
\usepackage{epstopdf}
\usepackage{capt-of}
\usepackage{tabularx,colortbl}
\usepackage{mathtools}
\usepackage{array}
\usepackage{makecell}
\usepackage{stackengine}
\setstackEOL{\cr}
\usepackage[table]{xcolor}

\newcommand{\msun}{\,M$_{\mathrm{\odot}}$\,}
\newcommand{\SAX}{SAX J1808.4--3658}

\title[The binary evolution of SAX J1808.4--3658]{The binary evolution of SAX J1808.4--3658: Implications of an evolved donor star}

\author[A. J. Goodwin et. al.]{
A. J. Goodwin$^{1}$\thanks{E-mail: ajgoodwin.astro@gmail.com} and T. E. Woods$^{2}$\\ 
\\
$^{1}$School of Physics and Astronomy, Monash University, Clayton, Victoria, Australia, 3800\\
$^{2}$National Research Council of Canada, Herzberg Astronomy \& Astrophysics Research Centre,\\ 5071 West Saanich Road, Victoria, BC V9E 2E7, Canada\\
}

\date{Accepted 2020 April 28. Received 2020 April 15; in original form 2020 February 12}

\pubyear{2020}

\begin{document}
\label{firstpage}
\maketitle

\begin{abstract}

Observations of the accretion powered millisecond pulsar SAX J1808.4--3658 have revealed an interesting binary evolution, with the orbit of the system expanding at an accelerated rate. We use the recent finding that the accreted fuel in SAX J1808.4--3658 is hydrogen depleted to greatly refine models of the progenitor and prior evolution of the binary system. We constrain the initial mass of the companion star to 1.0--1.2\,M$_{\mathrm{\odot}}$, more massive than previous evolutionary studies of this system have assumed. We also infer the system must have undergone strongly non-conservative mass transfer in order to explain the observed orbital period changes. Following \citet{Jia2015}, we include mass loss due to the pulsar radiation pressure on the donor star, inducing an evaporative wind which is ejected at the inner Lagrangian point of the binary system. The resulting additional loss of angular momentum resolves the discrepancy between conservative mass transfer models and the observed orbital period derivative of this system. We also include a treatment of donor irradiation due to the accretion luminosity, and find this has a non-negligible effect on the evolution of the system. 


\end{abstract}




\section{Introduction}



Accretion powered millisecond pulsars (AMSPs) are rapidly spinning neutron stars in binary orbits that are thought to be the progenitors of radio millisecond pulsars \citep[e.g.,][]{vandenheuvel2006}. 
Pulsars are born with strong ($\approx10^{12}$\,G) magnetic fields, and their rotation periods spin down over time due to rotational loss of energy \citep[e.g.,][]{Bhattacharya1991}. The peculiar observation of fast-spinning radio pulsars with weak ($\approx10^{8}$\,G) magnetic fields in the 1980s motivated the theory that these are ``recycled" pulsars that have been spun up through accretion of mass from a binary companion star and now appear as radio millisecond pulsars \citep[e.g.,][]{Bhattacharya1991}. Further evidence for this theory was that a large fraction of radio millisecond pulsars were observed to be in binary systems \citep{Bhattacharya1991}. It was not until 1998 that this ``recycling scenario" theory was confirmed by the observation of an actively accreting millisecond pulsar (AMSP) in a binary system, \SAX\,
\citep{Chakrabarty1998,Wijnands1998}. Since this discovery, more than 17 other AMSP systems have been discovered \citep[see e.g.,][for a review]{Patruno2012}. Now, more than 20 years after the first AMSP was discovered, it is thought that millisecond pulsars go through an accretion phase, in which they are spun up and X-ray emission is observed. Then, when the mass transfer rate reduces, they switch on as rotation-powered radio millisecond pulsars. This theory is supported by the observation of swings between rotation and accretion power in the millisecond binary pulsar IGR J18245--2452 \citep{Papitto2013}, which provides direct evidence that a rotation powered radio millisecond pulsar can switch on during periods of low mass inflow in an AMSP system, demonstrating an evolutionary pathway for AMSPs to evolve into radio pulsars.

Numerical studies of the formation channels of AMSP systems have been successful in modelling their population as a whole, but individual systems are often insufficiently constrained to reliably determine their prior evolutionary history \citep[e.g.,][]{Podsiadlowski2002,Nelson2003}. Recent efforts have used state of the art models combined with accreting neutron star observations, specifically those that exhibit Type I (thermonuclear) X-ray bursts, to constrain system parameters such as the accreted fuel's composition, the accretion rate, and the neutron star mass and radius \citep{Goodwin2019MCMC}. These new constraints enable more precise, system dependent, models of the binary evolution to be obtained, potentially clearing the way to constraining the mass transfer history and efficiency of these systems.



\SAX\, goes into outburst approximately every 4 years \citep{Goodwin2019ATel,DelSanto2015,Markwardt2011,Markwardt2008ATel,Markwardt2002ATel}, and is the most well studied and observationally constrained AMSP of its kind. The orbital period evolution of \SAX\, has proven consistently puzzling, as its orbital period derivative was measured to be an order of magnitude larger than expected by conservative mass transfer models \citep{diSalvo2008}. 
The currently prevailing theory for the unusually rapid orbital evolution of \SAX\, is the 
radio-ejection model, in which a significant amount of matter is ejected from the inner Lagrangian point of the system during quiescence. In the radio-ejection model, the radiation pressure from the pulsar on the infalling accreted matter stops the matter at the inner Lagrangian point, and ejects it from the system \citep[e.g.][]{Burderi2002}. This evolutionary phase occurs during the pulsar's transition from an AMSP to a rotation-powered radio millisecond pulsar, in a semi-detached binary.

\SAX\, is one of the few individual AMSPs whose evolution has been extensively modelled, due to the large amount of observational data available. However, none of these evolutionary studies have accounted for the fact that the donor star in this system has been shown to be significantly evolved, as evidenced from model constraints on the accreted fuel composition during X-ray outbursts \citep{Goodwin2019MCMC, Johnston2018, Galloway2006}. An evolved donor increases the required initial mass of the donor star, since small ($<$0.7\,\msun) donors could not have depleted hydrogen in their cores within the Hubble time.

\citet{Tailo2018} simulated evolutionary tracks for \SAX\, using the stellar evolution code ATON 2.0 \citep{ATONcode}. They followed the binary evolution of the system beginning with an orbital period of $\approx$6.6\,hr and modelled evolution driven by angular momentum losses via magnetic braking, gravitational radiation, and mass loss. They accounted for effects of irradiation of the donor by the X-ray emission and the spin-down luminosity of the pulsar, but did not include evaporation in their models. They included a treatment of radio ejection, in which the system ejects mass lost from the donor, as well as its angular momentum, if the period of the system exceeds the critical period predicted. They start with a companion mass of 0.75\msun, NS mass of 1.33\msun and P=6.6\,hrs, assuming a NS radius of 10\,km. 

\citet{Chen2017} also simulated evolutionary tracks of \SAX\, using MESA, where the evolution of the binary system was driven by angular momentum losses from gravitational wave radiation, magnetic braking, and mass loss. They included evaporation driven by the spin-down luminosity of the pulsar and irradiation driven by the high X-ray luminosity during transient outburst phases. These processes induce a high wind-loss rate ($\sim10^{-9}$\,\msun\,yr$^{-1}$, eventually resulting in a detached binary system, motivating the authors to conclude that \SAX\, would evolve into a black widow-like system.

Both of these studies required that magnetic braking continues to be active once the companion star becomes fully convective, with \citet{Tailo2018} arguing that due to effects of irradiation on the internal structure of the companion star, it would not be completely convective even at its very small presently observed mass. These studies found that a mass loss rate of $\sim10^{-9}$\msun\,yr$^{-1}$ from the inner Lagrangian point is required in order to match the observed orbital period derivative of this system. Neither of the best evolutionary tracks of these studies included an evolved companion star, however, with approximately 50$\%$ helium at the surface at the current evolutionary phase of the system being required by the observed bursting behaviour. \citet{Tailo2018} do not discuss the H mass fraction at the surface of the donor star for their best model, but it is assumed to be somewhere between 0.7--0.75, and \citet{Chen2017} found that the current H mass fraction at the surface of the donor star was 0.68 in their best model. Neither of these H mass fractions support the evidence for a depleted H mass fraction (X$\approx0.57$) of the accreted fuel during X-ray outbursts found by \citet{Goodwin2019MCMC}, \citet{Johnston2018} as well as  \citet{Galloway2006}.


In particular, using a Markov Chain Monte Carlo method to match observations of accreting neutron stars in outburst with a semi-analytic ignition model, \citet{Goodwin2019MCMC} obtained constraints on the neutron star mass, radius, accretion rate, accreted fuel composition and distance to the AMSP SAX J1808.4--3658. They found that a present neutron star mass of approximately 1.6\msun (higher than the traditionally adopted value of 1.4\msun \citep[e.g.][]{Steiner2018} and a H fraction of approximately 0.57 for the accreted fuel were required in order to match burst observations. This result implies that not only has the neutron star gained mass through accretion, but that the companion star has undergone significant hydrogen-burning, placing a strong lower bound on its initial mass. In this work we calculate the binary evolution pathway of SAX J1808.4--3658 starting after the neutron star has formed, and taking into account the hydrogen depletion in the core of the donor star, in order to gain insight into the evolution of this system and its progenitor. 

In Section \ref{sec:methods} we describe the methods and physics used to model the evolutionary tracks of \SAX. In Section \ref{sec:results} we present our results, providing evolutionary tracks including the effects of evaporation and without the effects of evaporation. Finally in Section \ref{sec:discussion} we discuss the implications of our results and present our conclusions. 


\section{Methods}\label{sec:methods}

We calculated possible binary evolution pathways for SAX J1808.4--3658 using the Modules for Experiments in Stellar Astrophysics stellar evolution binary program version 9575 \citep[MESA;][]{Paxton2011, Paxton2013, Paxton2015, Paxton2019}. We explored a range of initial donor star masses (0.6--2.0\,\msun) and orbital separations (0.4--3\,d), with a fixed initial neutron star point mass of 1.4\,M$_{\odot}$. We initialised the donor star as a zero-age main sequence star with a hydrogen mass fraction of X=0.7 and metallicity of Z=0.02. We used the Ritter mass transfer scheme \citep{Ritter1988} and the Eggleton Roche Lobe Overflow scheme \citep{Eggleton1983} with Eddington limited accretion. 

The rate at which angular momentum is lost from the system is dependent on gravitational radiation, mass loss, and magnetic braking, with the total change in angular momentum given by

\begin{equation}
    \dot{J} = \dot{J}_{\mathrm{GR}} + \dot{J}_{\mathrm{ML}} + \dot{J}_{\mathrm{MB}}
\end{equation}

For the rate of angular momentum loss due to gravitational radiation, $\dot{J}_{\mathrm{GR}}$, we used the default MESA implementation of the Peters formulae \citep{Peters1964} assuming zero initial eccentricity, as described in \citet{Paxton2015},

\begin{equation}
\dot{J}_{\mathrm{GR}} = -\frac{32}{5c^5}\left(\frac{2\pi G}{P_\mathrm{orb}}\right)^{7/3}\frac{(M_{\mathrm{NS}} M_{\mathrm{d}})^2}{(M_{\mathrm{NS}} + M_{\mathrm{d}})^{2/3}}
\end{equation}

\noindent where c is the speed of light, $M_{\mathrm{d}}$ is the donor star mass, $M_{\mathrm{NS}}$ is the neutron star mass, G is the gravitational constant, and $P_{\mathrm{orb}}$ is the orbital period. 

For the angular momentum changes due to magnetic braking, $\dot{J}_{\mathrm{MB}}$, we used the default MESA prescription, following \citet{Rappaport1983},

\begin{equation}
\dot{J}_{\mathrm{MB}} = -3.8\times10^{-30}M_1 R_{\mathrm{\odot}}^4 \left(\frac{R_{\mathrm{d}}}{R_{\mathrm{\odot}}}\right)^{\gamma_{mb}} \omega^3
\end{equation}



\noindent where $\omega$ is the rotational angular frequency of the donor star, $R_{\mathrm{d}}$ is the radius of the donor star, and $\gamma_{mb}$ is the magnetic braking index, which we fixed at the default MESA value of 3.\footnote{Note that while \citet{Chen2017} write that a magnetic braking index of 4 was used in their calculations, we can only replicate their published results with the default MESA value, which is 3. The correct magnetic braking index is not very well constrained \cite[see][]{Rappaport1983}, so we simply adopt the MESA default value in this work.} 
We did not switch magnetic braking off once the donor star has sufficiently reduced in mass to (in isolation) become fully convective ($\sim 0.3M_{\odot}$). As discussed in \citet{Tailo2018}, due to the effects of irradiation on the internal structure of the donor star, it may not be fully convective at these very low masses, and magnetic braking could still be active in the system.

The final contribution to orbital angular momentum evolution is mass loss, which can cause different amounts of angular momentum loss depending on where the mass is ejected from the system. By default, MESA includes mass loss prescriptions for mass ejected via a fast isotropic wind from either star, or a circumbinary coplanar toroid. Here, we include mass loss due to the radiation pressure of the pulsar on the donor star, causing an evaporative wind to blow that is ejected at the inner Lagrangian point of the binary system. 
The default MESA mass transfer efficiency, $f_{\mathrm{mt}}$, is given by

\begin{equation}
    f_{\mathrm{mt}} = 1 - \alpha_{\mathrm{mt}} - \beta_{\mathrm{mt}} - \delta_{\mathrm{mt}}
\end{equation}

\noindent where $ \alpha_{\mathrm{mt}}$ is the efficiency of mass loss from the vicinity of the donor, $\beta_{\mathrm{mt}}$ from the accretor, and $\delta_{\mathrm{mt}}$ from the circumbinary coplanar toroid. We assumed there is no mass lost from the immediate vicinity of the donor or the circumbinary coplanar toroid, and chose an arbitrary $\beta_{\mathrm{mt}}$ = 0.5 for mass lost from the vicinity of the accretor. Upon testing, using different values of $\beta_{\mathrm{mt}}$ did not significantly affect the evolutionary tracks of our calculations. The results for the models we calculated using only the angular momentum losses outlined above are presented in Section \ref{sec:noevap}.

In addition to the standard MESA mass loss implementation, we also defined an evaporation efficiency, $f_{\mathrm{ev}}$, which is the efficiency by which mass is ejected via the evaporative wind at the inner Lagrangian point of the binary system. This is implemented via an extras routine in MESA, as the prescription is not included in the standard version of the software. As in \citet{Jia2015}, we assumed the pulsar radiation causes evaporation of the donor star during quiescence, when the mass transfer is temporarily interrupted, driving a wind which is given by \citep{vandenheuvel1988,Stevens1992}
\begin{equation}
    \dot{M}_{\mathrm{d,evap}} = -\frac{f_{\mathrm{ev}}}{2v_{\mathrm{d,esc}}^2}L_p\left(\frac{R_{\mathrm{d}}}{a}\right)^2
\end{equation}

\noindent where $v_{\mathrm{d,esc}}$ is the escape velocity of the donor star, $a$ is the orbital separation, and $L_p$ is the spin-down luminosity of the pulsar, $L_p = 4\pi^2 I\dot{P}/P^3$, where $I$ is the pulsar moment of inertia. This evaporative wind is then added to the total wind mass transfer from the donor star, which is ejected at the inner Lagrangian point, adding to the angular momentum losses of the system. For the pulsar properties, we assume the pulsar spin evolution follows the standard magnetic dipole radiation model described by \citet{Shapiro1983}, assuming $P_0=3$\,ms (and in our calculations $P_f=3.1$\,ms), $\dot{P}=1\times10^{-20}$\,ss$^{-1}$, $B\sim10^8$\,G, and $I=10^{45}$\,gcm$^{-2}$. The evolution of the pulsar spin period we adopt thus neglects spin-up due to accretion, however, on comparison of a model calculation with a faster initial spin period pf $P_0=2.4$\,ms (and $P_f=2.5$\,ms), the evolutionary tracks were very similar, and the best fit model remained the same.

The total angular momentum changes due to mass loss in our calculations are thus given by

\begin{equation}
    \dot{J}_{\mathrm{ML}} = (\beta_{\mathrm{mt}}\dot{M}_{\mathrm{d,RLOF}} + \dot{M_{\mathrm{d,w}}}) L_1 a^2 \omega
\end{equation}

\noindent where $\omega=2\pi/P$ and $L_1$ is the distance from the inner Lagrangian point to the center of mass, in units of the separation, $a$. As in \citet{Beer2007}, if $\frac{M_{\mathrm{NS}}}{M_{\mathrm{d}}} \leq 10.0$ then $L_1$ is given by \citet{Warner1976}

\begin{equation}
   L_1 = 0.5 + 0.227 \mathrm{log}\frac{M_{\mathrm{NS}}}{M_{\mathrm{d}}} 
\end{equation}

\noindent If $\frac{M_{\mathrm{NS}}}{M_{\mathrm{d}}} \geq 10.0$ then $L_1$ is given by \citet{Kopal1959}

\begin{equation}
    L_1 = \left|\frac{M_{\mathrm{d}}}{M_{\mathrm{d}}+ M_{\mathrm{NS}}} - (1.0 - w_K + \frac{w_K^2}{3.0} + \frac{w_K^3}{9.0})\right|
\end{equation}
    

\begin{equation}
     w_K = \left(\frac{1.0}{3.0(1+(M_{\mathrm{NS}}/M_{\mathrm{d}}))}\right)^{1/3}
\end{equation}
   
Thus, in our calculations, any mass that is released from the donor and not transferred to the accretor is ejected from the system with the specific angular momentum of the inner Lagrangian point. The results for the models we calculated including donor evaporation and mass ejection at the inner Lagrangian point are given in Section \ref{sec:withevap}. 

We also included irradiation of the donor star due to the X-ray accretion luminosity. We did not include irradiation due to heating caused by the pulsar luminosity, as this is typically orders of magnitude lower than the heating caused by accretion. We used the MESA accretion-powered irradiation prescription, which requires an accretor radius (11.2\,km \citep{Steiner2018}) and then calculates the X-ray luminosity as:

\begin{equation}
    L_X = \frac{\mathrm{G} M_{\mathrm{NS}} \dot{M}_{\mathrm{NS}}}{ R_{\mathrm{NS}}}
\end{equation}
 where $\dot{M}_{\mathrm{NS}}$ is the accretion rate onto the neutron star. 
 
 The irradiation flux incident on the companion is then calculated as:
 
 \begin{equation}
     F_{\mathrm{irr}} = \epsilon_{\mathrm{irr}}\frac{L_X}{4\pi a^2}
 \end{equation}
 where $a$ is the binary separation. 
 
 We chose to deposit the extra heating due to irradiation at a column depth of 10\,g\,cm$^{-2}$, at the very surface of the companion star. Finally, we limited the maximum irradiation flux to $3\times10^9$\,erg\,s$^{-1}$\,cm$^{-2}$, since there is an unknown irradiation efficiency of the accretion luminosity. It is likely the accretion disk could prevent some of the accretion flux being incident on the companion star, and most likely all of the accretion luminosity would not irradiate the companion. \citet{Tailo2018} assumed an irradiation efficiency in the range of 1$\%$--2$\%$. We arrive at a maximum irradiation flux of $3\times10^9$\,erg\,s$^{-1}$\,cm$^{-2}$ by assuming $M_{\mathrm{NS}} = 1.4$\,\msun, $R_{\mathrm{NS}}=11.2$\,km, the time-averaged long term accretion rate of the system, $\dot{M}_{\mathrm{NS}}=2.55\times10^{12}$\,
 msun/yr, $\epsilon_{\mathrm{irr}}=0.015$, and an average orbital separation of $a=1\times10^{11}$\,cm. The companion star is consistently irradiated at $3\times10^9$\,erg\,s$^{-1}$\,cm$^{-2}$ for the duration of accretion in our calculations.
 
 The results for the models we calculated including donor evaporation, mass ejection at the inner Lagrangian point, and irradiation are given in Section \ref{sec:withevapandirrad}. 
 
We used the default MESA binary timestep controls for all evolutionary tracks calculated. As a test, we reduced all of the `varcontrol' timestep parameters by a factor of 10 and observed no difference in the predicted evolutionary tracks for the smaller timesteps.

The observed orbital epheremeris and predicted system parameters for \SAX, compiled from more than 20 years of observations of the source, are listed in Table \ref{tab:observations}. We calculated a grid of models with a range of initial companion star masses and initial orbital periods, fixing the neutron star mass to 1.4\msun, and the initial composition of the companion star to be $X=0.7$, $Z=0.02$. We determined the preferred initial companion star mass, and minimum and maximum possible orbital periods as outlined in Section \ref{sec:results}, and varied the mass in increments of 0.05\msun, orbital period in increments of 0.05\,d, and evaporation efficiency in increments of 0.01.

\begin{table}
	\centering
	\caption{
	Observed parameters for \SAX}
	\label{tab:observations}
	\begin{tabular}{llll} 
		\hline
		Parameter & Value & Units & Ref. \\
		\hline
		$P_{\mathrm{orb}}$& 7249.1569$\pm$0.0001 & s & 1\\
		$\dot{P_{\mathrm{orb}}}$ & (1.6$\pm$0.7)$\times10^{-12}$ &s\,s$^{-1}$ & 2, 3, 4, 5 \\
		$M_{\mathrm{NS}}$ & 1.5$^{+0.6}_{-0.3}$ &\msun & 6 \\
		$M_{\mathrm{c}}$ & 0.05$^{+0.05}_{-0.03}$ &\msun & 7 \\ 
		$X_{\mathrm{c}}$ & 0.57$^{+0.13}_{-0.14}$& & 6 \\

		\hline
	\end{tabular}
	\small{\\\textbf{Ref.:} 1. \cite{Papitto2005}, 2. \cite{Hartman2008}, 3. \cite{diSalvo2008}, 4. \cite{Hartman2009}, 5. \cite{Bult2019b}, 6. \cite{Goodwin2019MCMC}, 7. \cite{Bildsten2001}}
\end{table}

\section{Results}
\label{sec:results}

\subsection{Minimum initial mass of companion star}

In order to set a strong lower bound on the minimum required mass of the companion star, we used MESA star to evolve a single star over the age of the Universe (which we assume is 14 billion years), to find the minimum mass star which can achieve a central hydrogen fraction of 0.57. We examined the central hydrogen fraction at the end of evolution for a range of initial masses, and found that a star with initial mass of 0.6\msun is the minimum mass that has a central hydrogen fraction of 0.57 within 14 billion years of evolution.

\subsection{Maximum and minimum initial orbital periods}
The maximum initial period is set by the bifurcation period, which is the maximum period at which a system could evolve into a close low mass X-ray binary with an ultra-short period within the Hubble time \citep{vanderSluys2005}. For a \SAX-like system the bifurcation period is approximately 3 d \citep{vanderSluys2005}. 

We determined the minimum initial orbital period by calculating the Roche Lobe radius of the system for an adopted minimum mass ratio of $q=0.5$ with $M_{\mathrm{d}} = 0.7$\msun and $M_{\mathrm{NS}} = 1.4$\msun. We used the Roche Lobe formula from \citet{Eggleton1983},

\begin{equation}
    \mathrm{rL} = \frac{0.49 q^{2/3}}{0.6 q^{2/3} + \mathrm{ln}(1+q^{1/3})}
\end{equation}

\noindent and then found the separation to be R$_{\mathrm{ZAMS}}$/rL, where  R$_{\mathrm{ZAMS}} = 6.234\times10^{10}$\,cm to infer the period. This minimum period is 0.4\,d.

\subsection{Evolution without donor evaporation}\label{sec:noevap}

We first calculated models (labelled ``A'') assuming conservative mass transfer (setting $\beta_{\mathrm{mt}}=0.5$) with no evaporative wind (i.e., $f_{\mathrm{ev}} = 0.0$) or donor irradiation.
A full list of the best-fitting Model A initial conditions and results is given in Table \ref{tab:modelA}.

\begin{table}
	\centering
	\caption{Model A results: evolution without donor evaporation}
	\label{tab:modelA}
	\begin{tabular}{lcc} 
		\hline
		Parameter & Value & Units\\
		\hline
		$M_{\mathrm{d,i}}$& 1.1 &\msun\\
		$M_{\mathrm{d,f}}$ & 0.02&\msun\\
		$M_{\mathrm{NS,i}}$ & 1.4 &\msun\\
		$M_{\mathrm{NS,f}}$ &1.94  &\msun\\ 
		$P_{\mathrm{i}}$ & 1.22&\,d \\
		$P_{\mathrm{f}}$ & 0.0834 &\,d \\
		$\dot{P_{\mathrm{f}}}$ & 5.07e-14 &\,s\,s$^{-1}$ \\
		$f_{\mathrm{ev}}$ & 0.0 & \\
		$X_{\mathrm{c,i}}$ & 0.7& \\
		$X_{\mathrm{c,f}}$ &0.52 & \\

		\hline
	\end{tabular}
\end{table}

\begin{figure*}[p]
  \sbox0{\begin{tabular}{@{}cc@{}}
    \includegraphics[width=240mm]{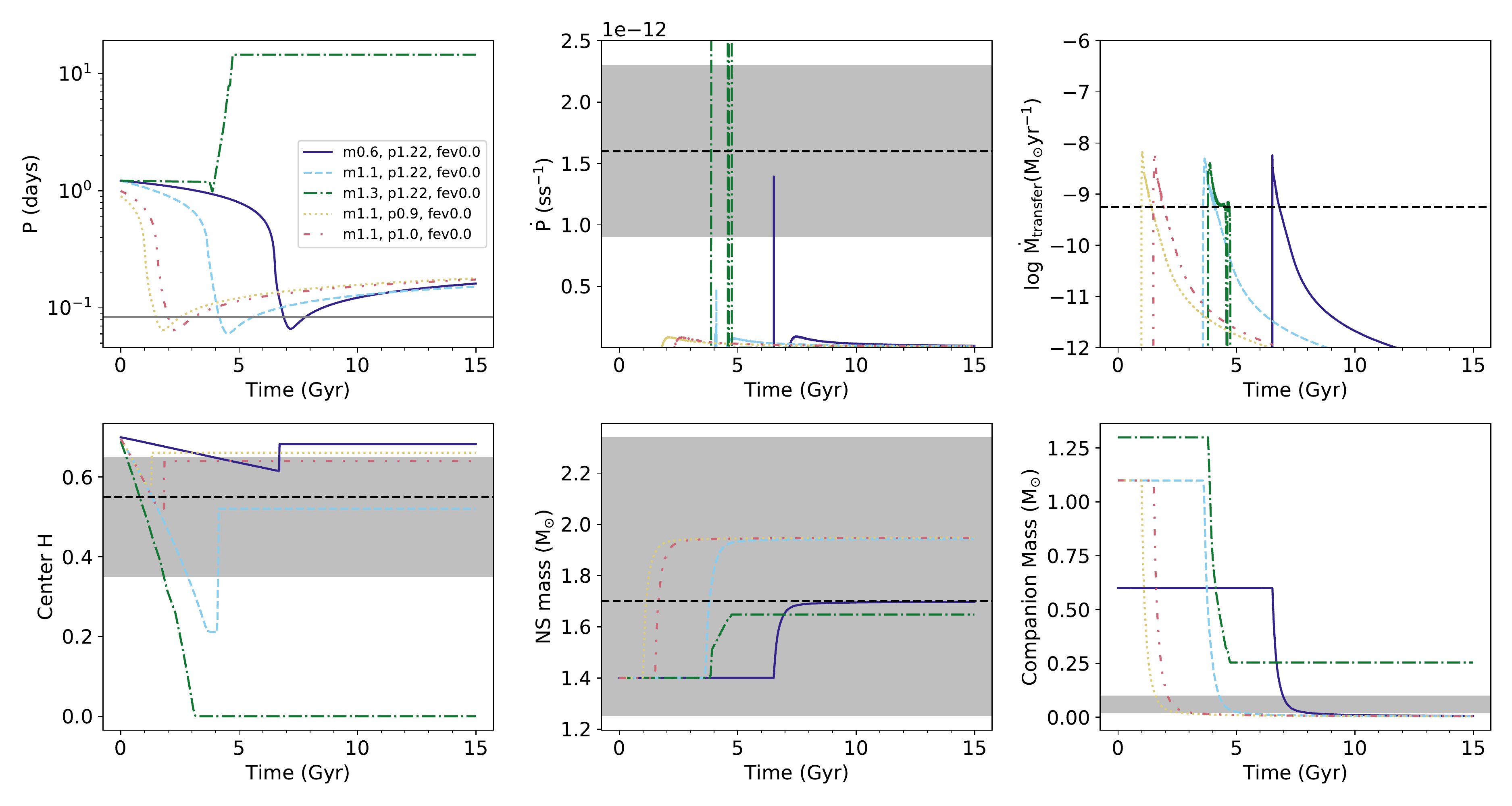}
  \end{tabular}}
  \rotatebox{90}{\begin{minipage}[c][\textwidth][c]{\wd0}
    \usebox0
    \captionof{figure}{Model A results: evolution without donor evaporation or irradiation ($f_{\mathrm{ev}}=0.0$) for a range of initial companion masses ($M_{\mathrm{d, i}}$) and orbital periods ($P_{\mathrm{i}}$). The different evolutionary tracks plotted are: $M_{\mathrm{d, i}} = 0.6$\msun, $P_{\mathrm{i}}=1.22$ d (purple solid line), $M_{\mathrm{d, i}} = 1.1$\msun, $P_{\mathrm{i}}=1.22$ d (blue dashed line), $M_{\mathrm{d, i}} = 1.3$\msun, $P_{\mathrm{i}}=1.22$ d (green dashed-dot line), $M_{\mathrm{d, i}} = 1.1$\msun, $P_{\mathrm{i}}=0.9$ d (orange dotted line), and $M_{\mathrm{d, i}} = 1.1$\msun, $P_{\mathrm{i}}=1.0$ d (red dashed-dot-dot line). All panels show the parameter evolution over time, with the individual panels showing: \textit{Top panels, left to right:} Orbital period, orbital period derivative, mass transfer rate onto the NS. \textit{Bottom panels, left to right:} Central hydrogen mass fraction of the donor star, NS mass, donor star mass. The dashed black line and grey regions indicate the current observed value of the parameter, and uncertainty, respectively.}
    \label{fig:nomasslossmodels}
  \end{minipage}}
\end{figure*}

We find that Model A is unable to reproduce the observed properties of \SAX, in particular the orbital period derivative. In Figure \ref{fig:nomasslossmodels}, it is clear that, at the current observed orbital period of 2.01 hours, the predicted orbital period derivative of the system is approximately two orders of magnitude lower than the observed value for all model combinations of initial companion masses and orbital periods that evolve to have a period as short as the current observed orbital period of the system. This finding is consistent with the earlier work of  \citet{diSalvo2008} and \cite{Burderi2009}, who showed that such a high orbital period derivative can not be accounted for by conservative mass transfer alone. As concluded by \citet{Hartman2008}, \cite{Hartman2009}, \cite{Patruno2017}, \cite{Chen2017}, and \cite{Tailo2018}, the rapid change in the orbit that is inducing such a high orbital period derivative could be caused by either a highly inefficient mass transfer scenario that this model does not account for, or the observed orbital period derivative could be a short term evolutionary phase caused by tidal dissipation and magnetic activity in the companion. This kind of short timescale orbital period evolution should have quasi-cyclic variability with timescales $\sim$10\,years. The latest outburst of \SAX\, \citep{Bult2019,Goodwin2019ATel} provided the 20th year of observations of the orbital period derivative, and has shown an interesting development. \citet{Bult2019b} measured a long-term orbital period derivative of $(1.6\pm0.7)\times10^{-12}$\,ss$^{-1}$, finding an interesting possible quasi-periodic variability in the orbit with a $\approx7$\,s amplitude around an expanding orbit, or a $\approx20$\,s amplitude modulation around a constant binary orbit. Additional monitoring of future outbursts would be necessary to differentiate between these scenarios. In the next section, we model the evolution assuming that the binary orbit is not constant, and is expanding at an accelerated rate.

\subsection{Evolution with donor evaporation}\label{sec:withevap}

We then evolved a grid of models with evaporation efficiencies ranging from 0.001--0.7, in which the evaporative wind is ejected at the inner Lagrangian point of the binary system with no irradiation of the donor star (labelled ``Model B"), as described in Section \ref{sec:methods}. The parameters for the model with the closest match to the observed values, as well as selected models with different initial orbital periods, companion masses, and evaporation efficiencies, are listed in Table \ref{tab:modelB} and plotted in Figures \ref{fig:massperiodmodels} and \ref{fig:evapefficiencymodels}.

\begin{table*}
	\centering
	\caption{Model B results for different initial companion masses, orbital periods, and evaporation efficiencies.}
	\label{tab:modelB}
	\begin{tabular}{lcccc} 
		\hline
		Parameter & Best fit & Higher $f_{\mathrm{ev}}$ & Lower $M_{\mathrm{c,f}}$ & Lower $P_{\mathrm{i}}$ \\
		\hline
		$M_{\mathrm{d,i}}$ (\msun) & 1.1 & 1.1 & 0.6 & 1.1 \\
		$M_{\mathrm{d,f}}$ (\msun) & 0.031 & 0.035 & 0.038 & 0.036 \\
		$M_{\mathrm{NS,i}}$ (\msun) & 1.4 & 1.4 & 1.4 & 1.4 \\
		$M_{\mathrm{NS,f}}$ (\msun) & 1.92 & 1.92 & 1.67 & 1.92 \\ 
		$P_{\mathrm{i}}$ (d) & 1.22 & 1.22 & 1.22 & 1.0  \\
		$P_{\mathrm{f}}$ (d) &  0.0832 & 0.0832 & 0.0836 & 0.0832 \\
		$\dot{P_{\mathrm{f}}}$ ($10^{-12}$\,s\,s$^{-1}$) & 1.65 & 3.27 & 1.41 & 1.29 \\
		$f_{\mathrm{ev}}$ &  0.2 & 0.4 & 0.2 & 0.2 \\
		$X_{\mathrm{c,i}}$ & 0.7 & 0.7 & 0.7 & 0.7 \\
		$X_{\mathrm{c,f}}$ & 0.52 & 0.52 & 0.68 & 0.64 \\

		\hline
	\end{tabular}
\end{table*}

The best fit model including only donor evaporation, with $M_{\mathrm{d,i}} = 1.1$\msun, $P_{\mathrm{i}} = 1.22$\,d, and $f_{\mathrm{ev}}$ = 0.2, matches all of the observed system parameters within their respective uncertainties. Due to the uncertainty of the measured orbital period derivative, a range of evaporation efficiencies (0.15--0.3) produce orbital period derivatives at the current observed orbital period of the system that match the observed change with time. Thus, we cannot constrain the evaporation efficiency with higher accuracy, but adopt $f_{\mathrm{ev}}\approx0.2$ as the ``best'' value for Model B. 

\subsection{Evolution with donor evaporation and irradiation}\label{sec:withevapandirrad}

\subsubsection{Retracing the evolution of \SAX}
\begin{figure*}
  \sbox0{\begin{tabular}{@{}cc@{}}
    \includegraphics[width=240mm]{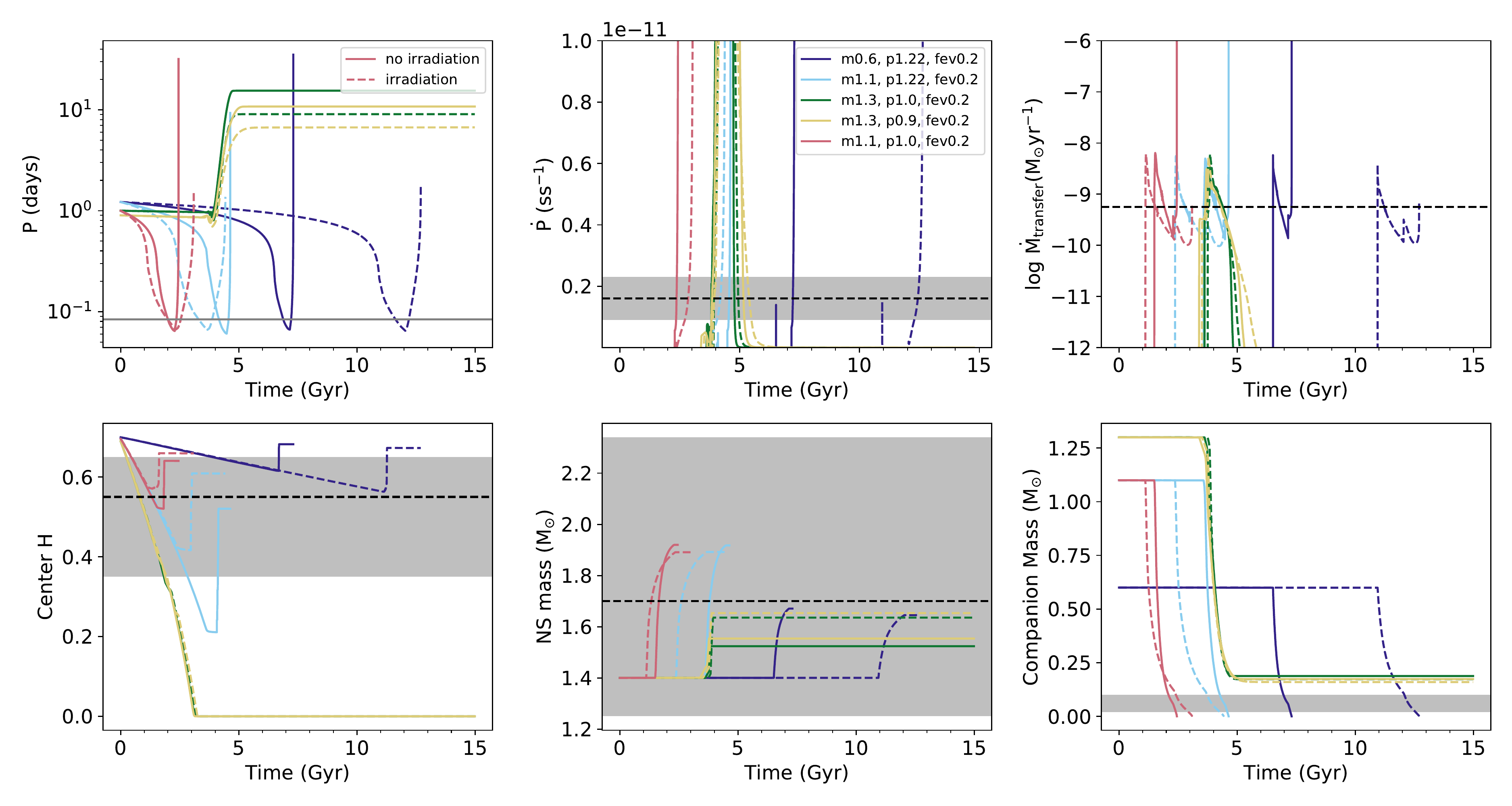}
  \end{tabular}}
  \rotatebox{90}{\begin{minipage}[c][\textwidth][c]{\wd0}
    \usebox0
    \captionof{figure}{Model B (solid line) and Model C (dashed line) results: Evolution with donor evaporation and irradiation for a range of initial companion masses ($M_{\mathrm{d, i}}$) and orbital periods ($P_{\mathrm{i}}$), with fixed evaporation efficiency ($f_{\mathrm{ev}}$). The different evolutionary tracks plotted are: $M_{\mathrm{d, i}} = 0.6$\msun, $P_{\mathrm{i}}=1.22$ d, $f_{\mathrm{ev}} = 0.2$ (purple), $M_{\mathrm{d, i}} = 1.1$\msun, $P_{\mathrm{i}}=1.22$ d, $f_{\mathrm{ev}} = 0.2$ (blue), $M_{\mathrm{d, i}} = 1.3$\msun, $P_{\mathrm{i}}=1.0$ d, $f_{\mathrm{ev}} = 0.2$ (green), $M_{\mathrm{d, i}} = 1.3$\msun, $P_{\mathrm{i}}=0.9$ d, $f_{\mathrm{ev}} = 0.2$ (orange), and $M_{\mathrm{d, i}} = 1.1$\msun, $P_{\mathrm{i}}=1.0$ d, $f_{\mathrm{ev}} = 0.2$ (red). All panels show the parameter evolution over time, with the individual panels showing: \textit{Top panels, left to right:} Orbital period, orbital period derivative, mass transfer rate onto the NS. \textit{Bottom panels, left to right:} Central hydrogen mass fraction of the donor star, NS mass, donor star mass. The dashed black line and grey regions indicate the current observed value of the parameter, and uncertainty, respectively.}
    \label{fig:massperiodmodels}
  \end{minipage}}
  \end{figure*}

\begin{figure*}
  \sbox0{\begin{tabular}{@{}cc@{}}
    \includegraphics[width=240mm]{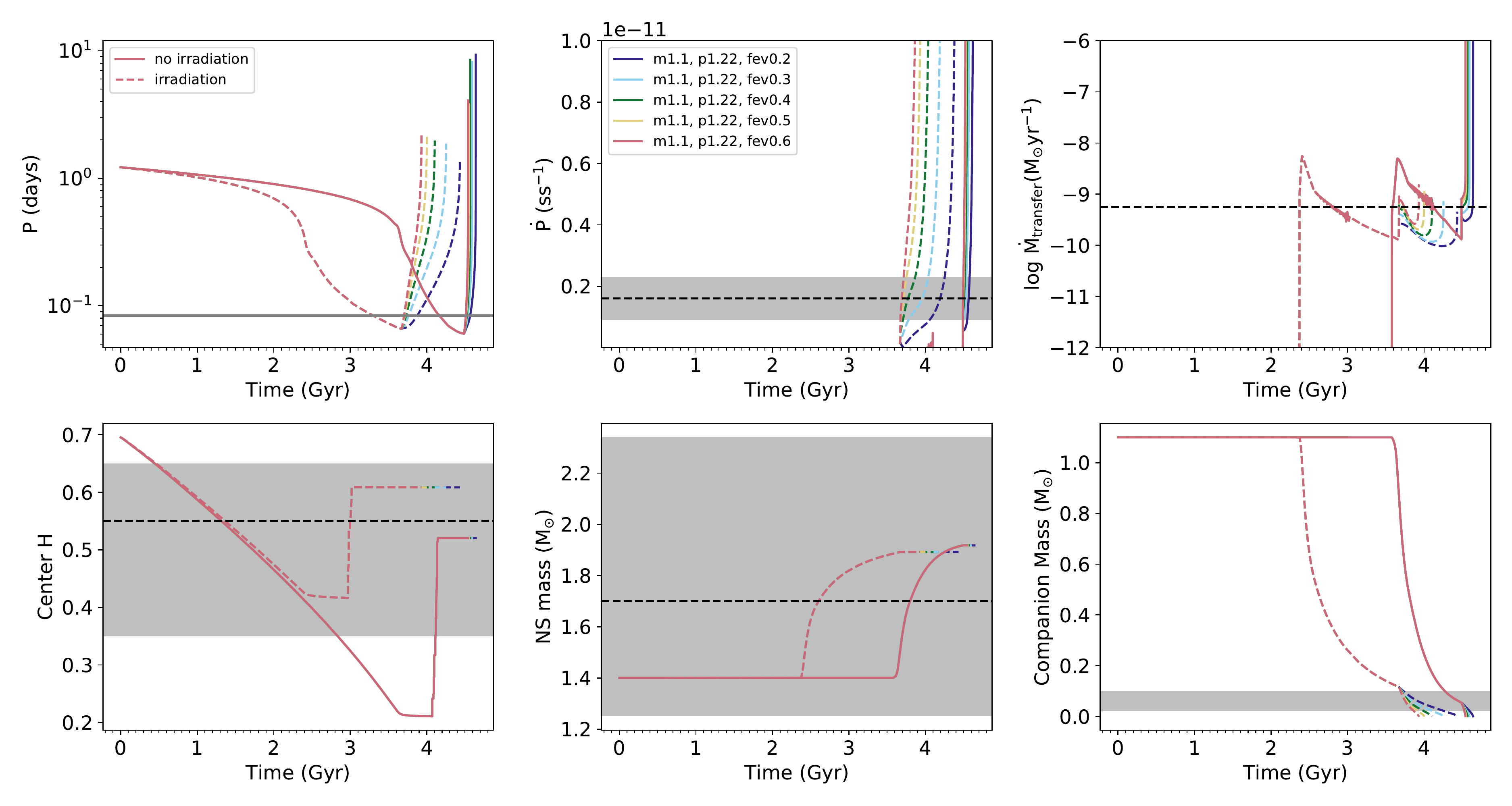}
  \end{tabular}}
  \rotatebox{90}{\begin{minipage}[c][\textwidth][c]{\wd0}
    \usebox0
    \captionof{figure}{Model B (solid line) and Model C (dashed line) results: Evolution with donor evaporation and irradiation for a range of initial companion masses ($M_{\mathrm{d, i}}$) and orbital periods ($P_{\mathrm{i}}$), with fixed evaporation efficiency ($f_{\mathrm{ev}}$). The different evolutionary tracks plotted are: $M_{\mathrm{d, i}} = 1.1$\msun, $P_{\mathrm{i}}=1.22$ d, $f_{\mathrm{ev}} = 0.2$ (purple), $M_{\mathrm{d, i}} = 1.1$\msun, $P_{\mathrm{i}}=1.22$ d, $f_{\mathrm{ev}} = 0.3$ (blue), $M_{\mathrm{d, i}} = 1.1$\msun, $P_{\mathrm{i}}=1.22$ d, $f_{\mathrm{ev}} = 0.4$ (green), $M_{\mathrm{d, i}} = 1.1$\msun, $P_{\mathrm{i}}=1.22$ d, $f_{\mathrm{ev}} = 0.5$ (orange), and $M_{\mathrm{d, i}} = 1.1$\msun, $P_{\mathrm{i}}=1.22$ d, $f_{\mathrm{ev}} = 0.6$ (red). All panels show the parameter evolution over time, with the individual panels showing: \textit{Top panels, left to right:} Orbital period, orbital period derivative, mass transfer rate onto the NS. \textit{Bottom panels, left to right:} Central hydrogen mass fraction of the donor star, NS mass, donor star mass. The dashed black line and grey regions indicate the current observed value of the parameter, and uncertainty, respectively.}
    \label{fig:evapefficiencymodels}
  \end{minipage}}
\end{figure*}

Finally, we evolved a grid of models with evaporation efficiencies ranging from 0.001--0.7, in which the evaporative wind is ejected at the inner Lagrangian point of the binary system and including the effects of irradiation of the donor star due to the accretion luminosity (labelled ``Model C"), as described in Section \ref{sec:methods}. The parameters for the model with the closest match to the observed values, as well as selected models with different initial orbital periods, companion masses, and evaporation efficiencies, are listed in Table \ref{tab:modelC} and plotted in Figures \ref{fig:massperiodmodels} and \ref{fig:evapefficiencymodels}. 

\begin{table*}
	\centering
	\caption{Model C results for different initial companion masses, orbital periods, and evaporation efficiencies.}
	\label{tab:modelC}
	\begin{tabular}{lcccc} 
		\hline
		Parameter & Best fit & Lower $f_{\mathrm{ev}}$ & Lower $M_{\mathrm{c,f}}$ & Lower $P_{\mathrm{i}}$ \\
		\hline
		$M_{\mathrm{d,i}}$ (\msun) & 1.1 & 1.1 & 0.6 & 1.1 \\
		$M_{\mathrm{d,f}}$ (\msun) & 0.086 & 0.068 & 0.173 & 0.0920 \\
		$M_{\mathrm{NS,i}}$ (\msun) & 1.4 & 1.4 & 1.4 & 1.4 \\
		$M_{\mathrm{NS,f}}$ (\msun) & 1.89 & 1.89 & 1.61 & 1.89 \\ 
		$P_{\mathrm{i}}$ (d) & 1.22 & 1.22 & 1.22 & 1.0  \\
		$P_{\mathrm{f}}$ (d) &  0.0867 & 0.0830 & 0.0836 & 0.0831 \\
		$\dot{P_{\mathrm{f}}}$ ($10^{-12}$\,s\,s$^{-1}$) & 1.73 & 0.492 & -0.175 & 1.59 \\
		$f_{\mathrm{ev}}$ &  0.5 & 0.2 & 0.5 & 0.5 \\
		$X_{\mathrm{c,i}}$ & 0.7 & 0.7 & 0.7 & 0.7 \\
		$X_{\mathrm{c,f}}$ & 0.42 & 0.42 & 0.57 & 0.57 \\

		\hline
	\end{tabular}
\end{table*}

When including the effect of irradiation of the donor star on the binary evolution, we require a much higher evaporation efficiency ($f_{\mathrm{ev}} = 0.5$) in order to match the observed orbital period derivative, for the model with $M_{\mathrm{d, i}} = 1.1$\msun, $P_{\mathrm{i}}=1.22$\,d. For Model C, including irradiation we find an equally good fit of the observed parameters for a shorter initial orbital period of $P_{\mathrm{i}}=1.0$\,d, but not for a smaller initial companion mass. Similarly to Model B, due to the uncertainty of the measured orbital period derivative, a range of evaporation efficiencies (0.4--0.6) produce orbital period derivatives at the current observed orbital period of the system that match the observed orbital period derivative. Thus, we cannot constrain the evaporation efficiency with higher accuracy, but adopt $f_{\mathrm{ev}}\approx0.5$ as the ``best'' value for Model C. 

\subsubsection{The effect of evaporation and irradiation feedback on the donor star evolution}

\begin{figure*}
	\includegraphics[width=\columnwidth]{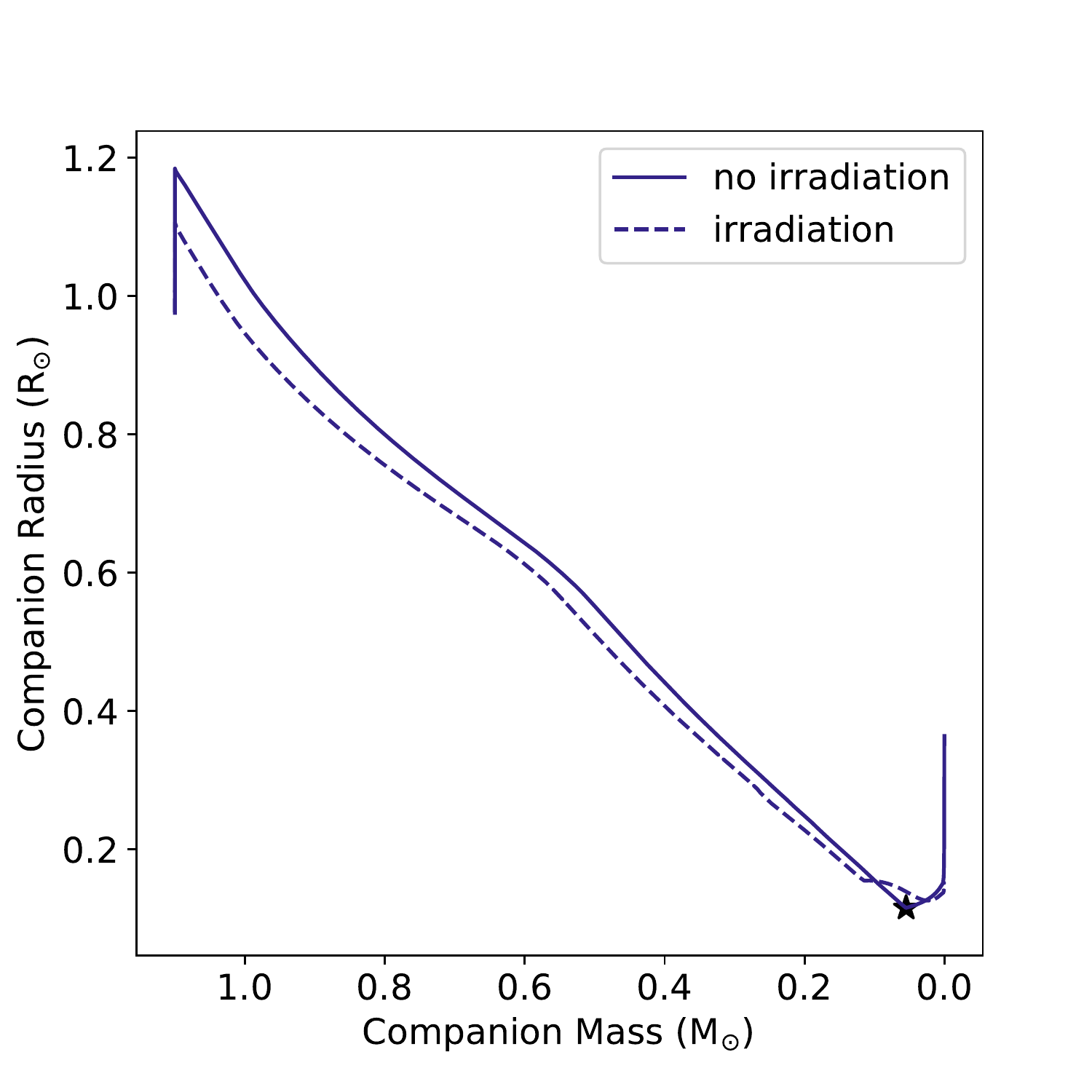}
	\includegraphics[width=\columnwidth]{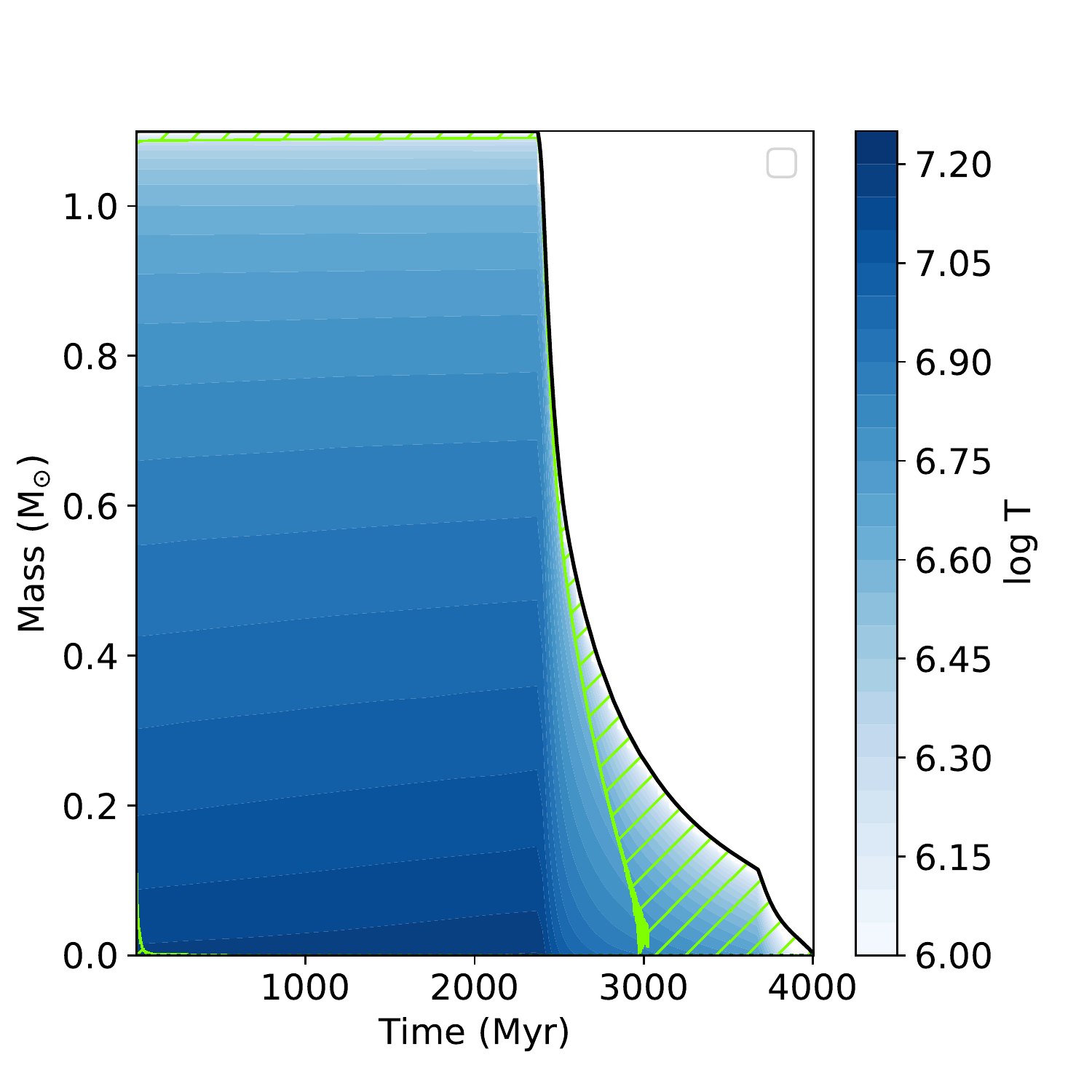}
    \caption{\textit{Left:} The evolution of the mass and radius of the companion star for Model B (no irradiation, solid line) and Model C (including irradiation, dashed line) for a model with $M_{\mathrm{d, i}} = 1.1$\msun, $P_{\mathrm{i}}=1.22$, $f_{\mathrm{ev}} = 0.5$. The black star indicates when evaporation begins, and it is clear that this evaporative feedback causes the companion star to expand. \textit{Right:} Kippenhahn diagram of the evolution of the internal structure of the donor star for a Model C calculation with $M_{\mathrm{d, i}} = 1.1$\msun, $P_{\mathrm{i}}=1.22$, $f_{\mathrm{ev}} = 0.5$. Green hatching indicates convective regions and blue shading is temperature (K). }
    \label{fig:donormassradius}
\end{figure*}

We explored the effect of evaporation and irradiation feedback on the evolution of the mass and radius of the donor star, and found that when the evaporative wind commences, the donor star immediately begins to expand while continuing to lose mass through Roche Lobe overflow. The evolution of the mass and radius of the donor star for Model B and Model C is plotted in Figure \ref{fig:donormassradius}, where the black star indicates the commencement of the evaporative wind. This rapid expansion confirms that due to the effects of the pulsar irradiation on the donor star, the surface of the star heats up and the star expands to become a very fluffy, low mass ``brown dwarf''. Interestingly, for Model C, in which we include the effects of donor irradiation by the accretion luminosity, the donor star does not expand nearly as much as Model B, in which we only include evaporation feedback.  

Figure \ref{fig:donormassradius} shows the evolution of the internal structure of the donor star as a Kippenhahn diagram (right panel). At $\mathrm{t}\approx3000$\,Myr, we see the convective envelope of the donor star reach all the way to the core of the star, and it becomes fully convective.

\section{Discussion}
\label{sec:discussion}
We modelled the binary evolution of \SAX, taking into account evidence that the donor star is significantly evolved, and the implications on the initial mass of the donor. We found the most likely progenitor of this system is a 1.1\msun companion star with an initial 1.0--1.22\,day orbital period. In order to match the observed orbital period derivative, our model requires that a significant amount of mass is ejected from the inner Lagrangian point of the binary system, by an evaporative wind caused by radiation pressure from the pulsar evaporating the donor star during quiescence. We also explored the effects of donor irradiation due to the accretion luminosity and found this has a non-negligible effect on the evolution of the system, particularly the donor star. These findings agree with previous calculations by \citet{Chen2017} and \citet{Tailo2018}.

In particular, our model requires a larger initial mass and orbital period than both \citet{Chen2017} and \citet{Tailo2018} in order to match the observed composition constraints of the accreted fuel. This requirement has implications for the progenitor systems of low mass X-ray binaries, and requires that the donor star in \SAX\, has undergone significant mass loss during the evolution of the system, and more than previously suspected.

Interestingly, in order to match the current observed system parameters, our model requires that magnetic braking remains active throughout the entire evolution of the system, even after the companion star becomes fully convective. \citet{Tailo2018} found that, due to the effects of irradiation, the companion star did not become fully convective, thus justifying keeping magnetic braking active. However, we found that even including irradiation of the donor star did not prevent the companion star from becoming fully convective. There are two points to consider in this regard: Firstly, our MESA models may not be appropriately evaluating stability against convection for the conditions in the irradiated donor, and the donor should actually still be at least partially radiative at its present mass. Secondly, the donor star in \SAX\, could in fact still have a strong magnetic field despite being fully convective, and therefore magnetic braking continues unabated. There have been some observations of fully convective dwarf stars with evidence for strong, stable magnetic fields \citep[e.g.,][]{Morin2008,Shulyak2017}. On theoretical grounds, some have proposed that intense activity and strong magnetism in fully convective dwarfs could be due to non-solar dynamo processes, concluding that fully convective stars could have significant magnetic fields that are generated in lower convection zones, and differential rotation contributes very little to the magnetic field \citep[e.g.][]{Durney1993}. 





In our best fit model (Model C), it takes $3.73\times10^{9}$\,years for the system to evolve from an orbital period of 1.22\,d with a companion star of 1.1\msun to the current observed period of 2.01\,hours with a 0.08\msun companion star. This model requires an evaporation efficiency of $f_{\mathrm{ev}}=0.5$. Since the initial primary must have been massive, and evolved very quickly through a common envelope (to bring it in to a short orbit) and a supernova, the timescale we find for the subsequent evolution of the donor is approximately the total present system lifetime, $\sim$3.7 billion years. 

As noted by numerous authors \citep[e.g.][]{Chakrabarty1998}, the binary parameters of \SAX\, are reminiscent of a black widow millisecond radio pulsar, which are known to ablate their companions and have very low mass companion stars. This work confirms that \SAX\, could indeed be a hidden black widow pulsar, since we require that \SAX\, switches on as a radio pulsar during quiescence in order to evaporate the companion star and eject mass at the inner Lagrangian point of the system. 

Optical observations of the 1998 and 2005 outbursts of \SAX\, revealed an optical i-band excess \citep{Greenhill2006,Wang2001}, which could be indicative of a circumbinary disk in this system. Future work could look into constraining theoretically the possibility of the existence of such a disk, and if the mass ejected at the inner Lagrangian point is sufficient to create it. 

\section{Conclusion}
We have demonstrated the necessity of taking into account that \SAX\, has an evolved donor star when modelling the evolution of the binary system. Our model requires an initial companion mass of 1.1\msun and orbital period of 1.0--1.22\,d in order to match the current observed system parameters (including for the first time the donor's measured hydrogen abundance) for an initial solar composition of the companion star and a 1.4\msun neutron star. As previous authors have found, conservative mass transfer models do not reproduce the observed orbital period derivative of the system. In this work, we find that including the effects of pulsar evaporation and irradiation on the donor star, ejecting the evaporative wind at the inner Lagrangian point of the binary system provides sufficient angular momentum losses to match the current observed rate of change of the orbital period. We emphasize that the effects of both irradiation of the donor star due to the accretion luminosity and evaporation of the donor star due to the pulsar luminosity are non-negligible, and both are a key component in describing the binary evolution of \SAX. 

\section*{Acknowledgements}

We thank Prof Alexander Heger and A/Prof Duncan Galloway for helpful comments and discussions. AG acknowledges support by an Australian Research Training Program scholarship. TEW acknowledges support from the NRC-Canada Plaskett fellowship. 




\bibliographystyle{mnras}
\bibliography{bibfile.bib}





\bsp	
\label{lastpage}
\end{document}